# Spin Pumping of an Easy-Plane Antiferromagnet Enhanced by Dzyaloshinskii-Moriya Interaction

Hailong Wang[1], Yuxuan Xiao[1,†], Mingda Guo[2,†], Eric Lee-Wong[3], Gerald Q. Yan[3], Ran Cheng[4,2], Chunhui Rita Du[3,1]

[1]Center for Memory and Recording Research, University of California, San Diego, La Jolla, California 92093
[2]Department of Physics and Astronomy, University of California, Riverside, California 92521
[3]Department of Physics, University of California, San Diego, La Jolla, California 92093
[4]Department of Electrical and Computer Engineering, University of California, Riverside, California 92521

[†]These authors contributed equally to this work.

Recently, antiferromagnets have received revived interest due to their significant potential for developing next-generation ultrafast magnetic storage. Here we report dc spin pumping by the acoustic resonant mode in a canted easy-plane antiferromagnet $\alpha$-$Fe_2O_3$ enabled by the Dzyaloshinskii-Moriya interaction. Systematic angle and frequency dependent measurements demonstrate that the observed spin pumping signals arise from resonance-induced spin injection and inverse spin Hall effect in $\alpha$-$Fe_2O_3$/metal heterostructures, mimicking the behavior of spin pumping in conventional ferromagnet/nonmagnet systems. The pure spin current nature is further corroborated by reversal of the polarity of spin pumping signals when the spin detector is switched from platinum to tungsten which has an opposite sign of the spin Hall angle. Our results highlight the potential opportunities offered by the low-frequency acoustic resonant mode in canted easy-plane antiferromagnets for developing next-generation, functional spintronic devices.

Over the past decade, magnetic resonance driven spin pumping has been demonstrated to be a powerful experimental technique for detecting pure spin currents in a range of magnet/nonmagnet heterostructures. Originally demonstrated in ferromagnetic metals and alloys,[1-10] this measurement platform has been quickly extended to magnetic insulators with reduced dissipation,[11-15] enabling a plethora of functional magnonic devices.[16-18] More recently, with the development of high-frequency microwave technologies and emerging interest in antiferromagnetic (AF) spintronics, sub-terahertz spin pumping has been demonstrated in antiferromagnets, such as $Cr_2O_3$[19] and $MnF_2$.[20] In contrast to (ferro)ferrimagnets, antiferromagnets promise to serve as a transformative platform enabling high-density memory, ultrafast data processing speeds, and robust strategies against malignant perturbations to develop next-generation spin-based information technologies.[21-26] Many of these advantages result from the ultrafast spin dynamics and vanishing net magnetic moment of the Néel order, which opens new opportunities for outperforming their ferromagnetic counterparts. A broad family of AF materials have been intensively explored at the frontier of spintronic research.[27-29]

Despite the growing availability of antiferromagnets with exceptional functionalities, AF spin pumping remains a formidable challenge at the current state of the art. The major difficulty derives from the exchange-amplified magnon gaps in typical uniaxial antiferromagnets,[19,20,30] which usually require (sub)terahertz frequency microwave capabilities that are not available in most commercially available electron paramagnetic resonance instruments. On the other hand, easy-plane antiferromagnets, which have smaller magnon gaps and accessible resonant frequencies, are believed to be not applicable for spin pumping because the pertaining spin dynamics is linearly polarized—a fundamental distinction from the uniaxial antiferromagnets. In this Letter, we challenge this traditional picture and report robust gigahertz (GHz) spin pumping from the canted magnetization in easy-plane antiferromagnet $\alpha$-$Fe_2O_3$.[31-34] Theoretical analysis reveals the critical role of the Dzyaloshinskii-Moriya (DM) interaction which enhances the spin pumping efficiency by orders of magnitude through modifying the ellipticity of the precessional motion of the canted magnetic moment. Our results highlight the previously-ignored opportunities offered by easy-plane antiferromagnets for developing next-generation, multifunctional spintronic devices.[35-41]

We first briefly review the pertinent magnetic properties of $\alpha$-$Fe_2O_3$. $\alpha$-$Fe_2O_3$ is an insulating antiferromagnet with a corundum crystal structure, as shown in Fig. 1(a). Above the Morin transition temperature, the $Fe^{3+}$ magnetic moments stay in the (0001)-plane and stack antiferromagnetically along the $c$-axis. The magnetic moment of the two AF spin sublattices are slightly canted by the DM interaction, leading to a weak ferromagnetic moment.[31-34] While the DM interaction only introduces a minor change to the ground state, it drastically changes the spin dynamics of the acoustic (low-frequency) resonant mode, as illustrated in Fig. 1(b). Suppose that an external magnetic field $H$ is applied along the $y$-axis in the magnetic easy plane of $\alpha$-$Fe_2O_3$. In absence of the DM interaction, the two sublattices magnetic moments $\boldsymbol{m}_1$ and $\boldsymbol{m}_2$ precess elliptically in a right-handed chirality, where the major (minor) axes of the trajectories are along the $y$($z$)-axis and angles have been exaggerated for visual clarity. Consequently, unless $H$ becomes extremely strong to induce appreciable canting of $\boldsymbol{m}_1$ and $\boldsymbol{m}_2$ along the $y$-axis, the total magnetic moment $\boldsymbol{m} = \boldsymbol{m}_1 + \boldsymbol{m}_2$ will exhibit an almost linear oscillation along the $z$ direction, as illustrated in the top panel of Fig. 1(b). Correspondingly, the oscillation of $\boldsymbol{m}(t)$ can only give rise to an ac spin pumping while the dc component is negligible. Moreover, an $x$-polarized microwave magnetic field is not able to drive the resonance of $\boldsymbol{m}(t)$ because of the polarization mismatch.

This is a universal property of the conventional easy-plane antiferromagnets, such as NiO and MnO. When the DM field $H_D$ is involved, however, $\boldsymbol{m}_1$ and $\boldsymbol{m}_2$ will be canted towards the $y$-axis at equilibrium and precess elliptically around the canted directions, resulting in an elliptically circulating $\boldsymbol{m}(t)$ around the $y$-axis, as illustrated in the bottom panel of Fig. 1(b). Because the elliptical trajectory of the canted magnetic moment now has a finite projection along $x$-axis direction, microwave magnetic field polarized along the Néel vector can effectively drive the acoustic resonant mode of $\alpha$-Fe$_2$O$_3$, hence a dc spin pumping. As $H_D$ increases, the precession of $\boldsymbol{m}(t)$ becomes more circular, leading to a significantly enhanced dc spin pumping efficiency in comparison with the conventional easy-plane antiferromagnets.

Our spin pumping measurements use 0.5-mm-thick commercially available $\alpha$-Fe$_2$O$_3$ crystals (see Supplementary Information Note 1 for details). Figure 1(c) shows a temperature dependence of the magnetization of an $\alpha$-Fe$_2$O$_3$ crystal with an external magnetic field $H$ of 1000 Oe applied in the (0001)-plane. The measured magnetization curve exhibits a sharp, step-like variation around the Morin transition temperature $T_M \sim 263$ K. When $T < T_M$, $\alpha$-Fe$_2$O$_3$ becomes a uniaxial antiferromagnet with a negligible ferromagnetic moment. When $T > T_M$, a weak canted magnetic moment of $\sim 2$ emu/cm$^3$ is observed,[32-34] which is also confirmed by the field-dependent magnetization curve shown in the inset. At room temperature, two magnetic resonance modes coexist in $\alpha$-Fe$_2$O$_3$. When an external magnetic field $H$ is applied in the (0001)-plane, the low-frequency (acoustic) mode described above has an eigenfrequency:[33,34]

$$\left(\frac{2\pi f_{\text{FMR}}}{\gamma}\right)^2 = H(H + H_D) + 2H_E H_A \quad (1)$$

where $\gamma$ is the gyromagnetic ratio, $f_{\text{FMR}}$ is resonant frequency, $H_E$ is the exchange field, $H_A$ is the effective field associated with the basal plane anisotropy, and $H_D$ is the DM field. The high-frequency mode (*a.k.a.* optical mode) corresponding to an out-of-plane oscillation of the Néel vector has a resonant frequency:[33,34]

$$\left(\frac{2\pi f_{\text{FMR}}}{\gamma}\right)^2 = H_D(H + H_D) + 2H_E H_{A'} \quad (2)$$

where $H_{A'}$ is the anisotropy field along the [0001]-axis (the hard axis). Figure 1(d) plots the expected resonant frequency $f_{\text{FMR}}$ of both modes as a function of $H$. In the magnetic field range of interest, $f_{\text{FMR}}$ of the low-frequency one lies in the low-GHz frequency regime,[33] which is accessible by the commercial microwave units. Our magnetic resonance and spin pumping measurements presented below will focus on the branch of the low-frequency mode of $\alpha$-Fe$_2$O$_3$.

To detect the resonance-driven spin currents, we use magnetron sputtering to deposit platinum (Pt) or tungsten (W) films on top of the $\alpha$-Fe$_2$O$_3$ crystals (see Supplementary Information Note 1 for details). The spin pumping measurements are performed using a microwave transmission line on the $\alpha$-Fe$_2$O$_3$/metal samples at room temperature (see Supplementary Information Note 2 for details). Figure 2(a) shows the schematic of the sample structure. A dc magnetic field $H$ is applied in the magnetic easy-plane ($x$-$y$ plane) and the spin pumping signal $V_{\text{SP}}$ is measured across the long axis ($x$-axis) of the sample. $\theta_H$ is defined as the relative angle between the in-plane magnetic field $H$ and the $y$-axis. At resonance condition, an oscillating Oersted field (linearly polarized along the $x$-axis) produced by microwave currents flowing a proximal transmission line compensates the Gilbert damping of $\alpha$-Fe$_2$O$_3$, then the canted magnetization can freely oscillate around the equilibrium position, transferring spin angular momenta to the conduction electrons of the neighboring metal layers through interfacial exchange coupling. The injected spin current $J_s$ is converted to a charge current: $J_c \propto \theta_{SH} J_s \sigma \cos \theta_H$ by the inverse spin Hall effect (ISHE),[1-4] where $\theta_{SH}$ is the spin Hall angle of the metal layer and $\sigma$ is the spin polarization parallel to the external magnetic field.

Figure 2(b) shows the normalized spin wave transmission spectrum measured as a function of magnetic field $H$ and microwave frequency $f$. At resonance condition, a significant amount of microwave power is absorbed by the $\alpha$-Fe$_2$O$_3$ crystal, leading to a reduced microwave transmission efficiency characterized by $S_{12}$. The dispersion curve of the observed spin wave mode shown in Fig. 2(b) agrees well with the theoretical prediction of the acoustic mode of $\alpha$-Fe$_2$O$_3$, by which $H_D$ and $2H_E H_A$ are fitted to be $(2.07 \pm 0.041) \times 10^4$ Oe and $(1.59 \pm 0.029) \times 10^7$ Oe$^2$, quantitatively agreeing with the values reported in previous studies.[33] Figures 2(c) and 2(d) show the field-dependent magnetic resonance and spin pumping spectra of an $\alpha$-Fe$_2$O$_3$/Pt (5nm) sample measured at $f = 14$ GHz and $\theta_H = 0°$. For spin pumping measurements, the input microwave power $P_{rf}$ of 1000 mW is modulated at a frequency of 161 Hz. To obtain a better signal-to-noise ratio, a preamplifier is used to amplify the electrical voltage signals with a gain of 500 (spin pumping data shown below were taken after amplification unless stated elsewise). It is clear that the electrical signals are observed at the magnetic fields satisfying resonance conditions. In addition, the measured spin pumping signals change sign when the external magnetic field is reversed from $\theta_H = 0°$ to $180°$, which is consistent with the theoretical model of ISHE. Figure 3(a) shows a series of spin pumping spectra of the $\alpha$-Fe$_2$O$_3$/Pt sample measured from 13 to 18 GHz with $\theta_H = 0°$. At all frequencies, the observed voltage signals show the expected sign reversal when the polarity of the external magnetic field changes. The variation of the absolute magnitude of the spin pumping signals results from the frequency-dependent microwave transmission efficiency. Figure 3(b) plots the magnetic field $H$ with the maximum spin pumping signals as a function of the microwave frequency, which essentially follows the acoustic resonance as described by Eq. (1). Figure 3(c) shows the spin pumping spectra of the $\alpha$-Fe$_2$O$_3$/Pt sample measured at microwave powers of 440, 1000, 1400, and 2000 mW with a frequency of 14 GHz. All the spectra have been offset by the magnetic resonance field $H_{res}$ for clarity. The inset shows the linear microwave power dependence of $V_{SP}$ measured at $\theta_H = 0°$.

To further confirm that the measured electrical voltages indeed result from ISHE, we employ tungsten (W) which has a negative spin Hall angle as an alternative spin detector.[42] Figure 4(a) shows the $V_{SP}$ vs $H$ spectra for two $\alpha$-Fe$_2$O$_3$/Pt and $\alpha$-Fe$_2$O$_3$/W samples measured at $P_{rf} = 1000$ mW and $\theta_H = 0°$. Notably, a robust voltage signal of 11.9 µV is observed in the $\alpha$-Fe$_2$O$_3$/W sample, showing the opposite sign in comparison with the $\alpha$-Fe$_2$O$_3$/Pt sample. When $H$ is rotated in the magnetic easy-plane of $\alpha$-Fe$_2$O$_3$, the canted magnetic moment remains essentially parallel to $H$ at all angles since the resonance field $H_{res}$ exceeds the basal plane anisotropy field. Figure 4(b) shows a series of $V_{SP}$ vs $H$ spectra for varying in-plane field angle $\theta_H$. When $H$ is rotated from $0°$ to $90°$, $V_{SP}$ observed in both samples gradually vanishes and approaches zero. Figure 4(c) summarizes the angular dependence of $V_{SP}$ for both $\alpha$-Fe$_2$O$_3$/Pt and $\alpha$-Fe$_2$O$_3$/W samples normalized by the maximum magnitude of spin pumping signals measured at $\theta_H = 0°$ (see Supplementary Information Note 2 for details). The clear cosinusoidal shape corroborates that the observed electrical voltages indeed arise from ISHE.

Next, we develop a theoretical model to quantitatively analyze the observed spin pumping in $\alpha$-Fe$_2$O$_3$/metal bilayers. According to the device geometry depicted in Fig. 2(a), a dc spin current density $I_s$ polarized along $y$-axis can be generated by an $x$-polarized microwave magnetic field $h_{rf}$ as follows (see Supplementary Information Note 3 for details):

$$I_s = \hbar g_r \langle \boldsymbol{m} \times \frac{d\boldsymbol{m}}{dt}\rangle = hg_r f \text{Im}[\chi_{xx}^* \chi_{zx}] h_{rf}^2 \tag{3}$$

where $h$ is the Planck constant, $f$ is the driving microwave frequency, $\chi_{xx}$ and $\chi_{zx}$ are two components of the dynamical susceptibility such that $m_i(t) = \text{Re}[\chi_{ij} h_{rf,j} e^{i2\pi ft}]$, and $g_r$ is the interfacial spin-mixing conductance.[43,44] In uniaxial antiferromagnets with circular polarization, the factor $\text{Im}[\chi_{xx}^* \chi_{zx}]$ is proportional to $\frac{H_A}{H^2 H_E}$ on resonance.[24,45] For easy-plane antiferromagnets with DM interaction, however, this factor becomes complicated due to the elliptical precession, so we resort to numerical calculations. Owing to the ISHE and the spin diffusion effect in the heavy metal, $I_s$ obtained from Eq. (3) converts into a spin pumping signal $V_{SP}$:[46]

$$V_{SP} = \xi \tilde{g}_r \frac{L\theta_{SH} \rho e}{2\pi} \frac{\lambda}{d_N} \tanh\left(\frac{d_N}{2\lambda}\right) h_{rf}^2 \tag{4}$$

where $e$ is the electron charge, $L$, $\theta_{SH}$, $\lambda$, $\rho$, and $d_N$ are the length, the spin Hall angle, the spin diffusion length, the resistivity, and the thickness of the heavy metal layer, respectively. The factor $\tilde{g}_r = \frac{g_r}{1+\frac{2\lambda e^2 \rho g_r}{h}\coth\frac{d_N}{\lambda}}$ is the effective spin-mixing conductance incorporating the spin backflow effect in the heavy metal.[47] Here we have defined $\xi = 2\pi f \text{Im}[\chi_{xx}^* \chi_{zx}]$ as the spin-pumping susceptibility to characterize the conversion efficiency of microwave power into dc spin current density. $\xi$ sensitively depends on the polarization status of $\boldsymbol{m}(t)$, which is determined by the DM field $H_D$, the driving frequency $f$, and the external magnetic field $H$. We calculate $\xi$ for $f = 14$ GHz and other parameters extracted from experiments,[48] where the Gilbert damping $\alpha = 4.6 \times 10^{-4}$ is obtained by comparing the measured linewidth with the numerical linewidth (see Supplementary Information Note 4 for details). In the unit of $\gamma/H_E$, Figure 5 plots $\xi$ as a function of $H$ for different strengths of $H_D$ with $H_D^{Max} = 2.07 \times 10^4$ Oe being the actual value we measured. Notably, the DM interaction significantly enlarges the peak value $\xi_p$ along with a reduction of the resonance field $H_{res}$. As plotted in the insets, when $H_D$ increases from zero to its actual value $H_D^{Max}$, the resonance field is reduced by 7 times while the peak value $\xi_p$ increases by almost 350 times. This drastic enhancement confirms the essential role of DM interaction underlying the observed dc spin pumping signals. Using $\theta_{SH} = 0.1$ $(-0.14)$,[14] $t_N = 5$ nm, $\lambda = 7.3$ $(2.1)$ nm,[14] and $\rho = 3.3 \times 10^{-7}$ $(2.5 \times 10^{-6})$ $\Omega$ m for the Pt (W) layer, and other relevant material parameters: $L = 4$ mm, $d_N = 5$ nm, and $h_{rf} = 0.2$ Oe at a frequency of 14 GHz (see Supplementary Information Note 5 for details), the effective spin-mixing conductance $\tilde{g}_r$ is estimated to be $7.1 \times 10^{20}$ m$^{-2}$ and $7.4 \times 10^{19}$ m$^{-2}$ for the $\alpha$-Fe$_2$O$_3$/Pt and $\alpha$-Fe$_2$O$_3$/W interfaces, respectively. These values are two orders of magnitude larger than those reported in (anti)ferromagnet/heavy-metal systems.[1,5,7,13,14,20,49] A recent study indicated that when the Néel vector contribution is suppressed, the spin-mixing conductance characterizing the spin pumping from $\boldsymbol{m}(t)$ could indeed be much larger than that for the Néel vector,[50] which is consistent with our results. However, it remains an open question as to why the spin-mixing conductance becomes significantly enhanced when the contribution of Néel vector precession vanishes. A possible reconciliation is that if we use the unit vector $\hat{\boldsymbol{m}}$, rather than $\boldsymbol{m}$ itself, to define spin pumping in Eq.(3), then the value of $\tilde{g}_r$ will shrink by a factor of $\left(\frac{H_D}{H_E}\right)^2$, yielding $5.5 \times 10^{17}$ m$^{-2}$ and $5.8 \times 10^{16}$ m$^{-2}$ for the $\alpha$-Fe$_2$O$_3$/Pt and $\alpha$-Fe$_2$O$_3$/W interfaces, respectively.

In conclusion, we have demonstrated robust GHz-frequency spin pumping of a canted easy-plane antiferromagnet $\alpha$-Fe$_2$O$_3$. By modifying the ellipticity of the resonant motion of the canted magnetic moment, the DM interaction could significantly enhance the spin pumping efficiency by orders of magnitude in spite of the detrimental role of hard-axis anisotropy, leading to robust inverse spin Hall signals measured in adjacent heavy metal layers. Our results suggest the exciting possibility of harnessing hard-axis antiferromagnets in developing next-generation functional magnetic devices.

*Note added*: Recently, a related paper by I. Boventer *et al.*, appeared.[51] They have also observed dc spin pumping signals in $\alpha$-Fe$_2$O$_3$/metal bilayers.

**Acknowledgements.** Authors would like to thank for fruitful discussions with Derek Reitz, Yaroslav Tserkovnyak, and Eric E. Fullerton. E. L.-W., G. Q. Y., and C. H. R. D. were supported by U. S. National Science Foundation under award ECCS-2029558 and DMR-2046227. C. H. R. D. acknowledges the support from Air Force Office of Scientific Research under award FA9550-20-1-0319 and its Young Investigator Program under award FA9550-21-1-0125. M. G. and R. C. were supported by the Air Force Office of Scientific Research under grant FA9550-19-1-0307.

**Figure captions:**

**Figure 1.** (a) Schematic of the crystal structure of $\alpha$-Fe$_2$O$_3$ (oxygen atoms not shown). The symbols and arrows indicate the Fe atoms and the spins associated with them. (b) Schematic illustration of the acoustic resonant mode in an easy-plane antiferromagnet with and without the DM field $H_D$. The external magnetic field $H$ is applied along the y-axis. (c) Temperature dependence of the magnetization of the $\alpha$-Fe$_2$O$_3$ crystal with an external magnetic field of 1000 Oe applied in the (0001) plane. The inset shows field-dependent magnetization of $\alpha$-Fe$_2$O$_3$ when $T$ = 295 K. (d) The expected resonant frequency $f_{FMR}$ of the low- and high-frequency modes as a function of external magnetic field. $H_D$, $2H_EH_A$ and $2H_EH_{A'}$ are set to be 22 kOe, 15 kOe$^2$, and 3200 kOe$^2$ in the calculations.[33,34]

**Figure 2.** (a) Schematic of dynamic spin injection and inverse spin Hall effect in an $\alpha$-Fe$_2$O$_3$/metal heterostructure. $\boldsymbol{m}_1$, $\boldsymbol{m}_2$, and $\boldsymbol{m}$ represent the magnetic moment of the two spin sublattices and the canted ferromagnetic moment of $\alpha$-Fe$_2$O$_3$, respectively. (b) Color-coded spin wave transmission spectrum $S_{12}$ measured as a function of external magnetic field $H$ and microwave frequency $f$. The white dashed line represents the fitting curve from which $H_D$ and $H_EH_A$ are extracted. (c) $S_{12}$ vs $H$ spectrum of $\alpha$-Fe$_2$O$_3$ measured at $f$ = 14 GHz and an input microwave power $P_{rf}$ = 1 mW. (d) $V_{SP}$ vs $H$ spectrum measured at $f$ = 14 GHz, $\theta_H$ = 0°, and $P_{rf}$ = 1000 mW.

**Figure 3.** (a) Spin pumping spectra of the $\alpha$-Fe$_2$O$_3$/Pt sample measured at different frequencies with $\theta_H$ = 0°. The curves are offset for clarity. (b) Peak position of the measured spin pumping signals as a function of the microwave frequency. A theoretical fitting (blue curve) using Eq. (1) agrees well with the experimental results (red points). (c) $V_{SP}$ vs ($H - H_{res}$) spectra of the $\alpha$-Fe$_2$O$_3$/Pt sample measured with input microwave power of 440, 1000, 1400, and 2000 mW (blue, green, black, and red curves, respectively) at $f$ = 14 GHz and $\theta_H$ = 0°. Inset: microwave power dependence of the measured spin pumping signals. The blue line is a linear fitting of the experimental data.

**Figure 4.** (a) $V_{SP}$ vs ($H - H_{res}$) spectra of the $\alpha$-Fe$_2$O$_3$/Pt sample (red curve) and the $\alpha$-Fe$_2$O$_3$/W sample (blue curve) measured at $f$ = 14 GHz and $\theta_H$ = 0°. (b) Spin pumping spectra of the $\alpha$-Fe$_2$O$_3$/Pt (red curves) and $\alpha$-Fe$_2$O$_3$/W (blue curves) measured at different in-plane field angle $\theta_H$. The curves are offset for clarity. (c) Angular dependence of $V_{SP}/|V_{SP}(0°)|$ for $\alpha$-Fe$_2$O$_3$/Pt and $\alpha$-Fe$_2$O$_3$/W. The red and blue curves show $\cos\theta_H$ and $-\cos\theta_H$, respectively.

**Figure 5.** Calculated spin-pumping susceptibility $\xi$ (scaled by unit $\gamma/H_E$) as a function of the external magnetic field $H$ for different DM fields, where $H_D^{Max}$ = 2.07 × 10$^4$ Oe is the actual value we measured. Insets: resonant field $H_{res}$ and resonant peak $\xi_p$ (normalized to its value at $H_D = 0$) as functions of $H_D/H_D^{Max}$. Parameters for the calculation are chosen from experiments[48] and the Gilbert damping coefficient $\alpha$ = 4.6 × 10$^{-4}$ is obtained by comparing the measured linewidth with the numerical values (see Supplementary Information Note 4 for details).

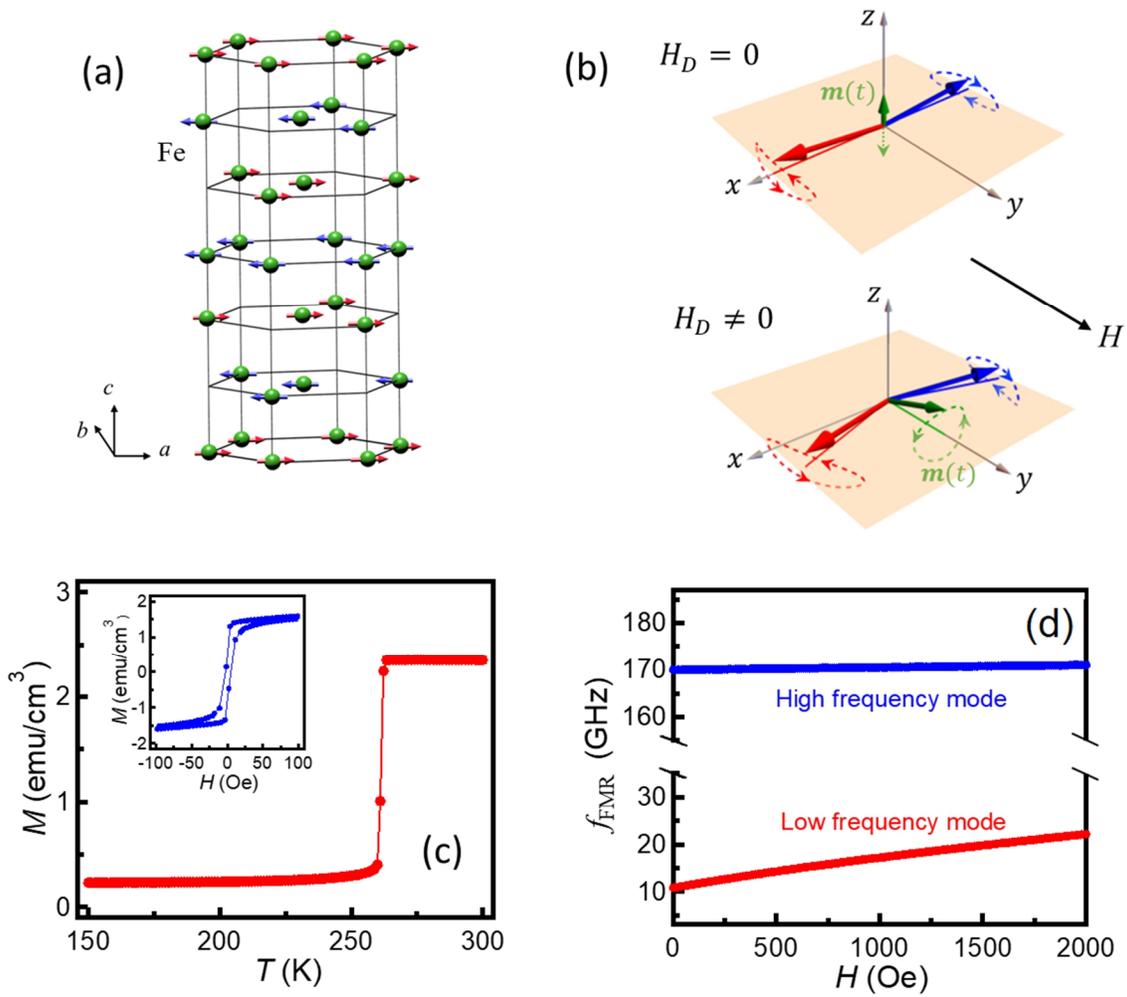

**Figure 1**

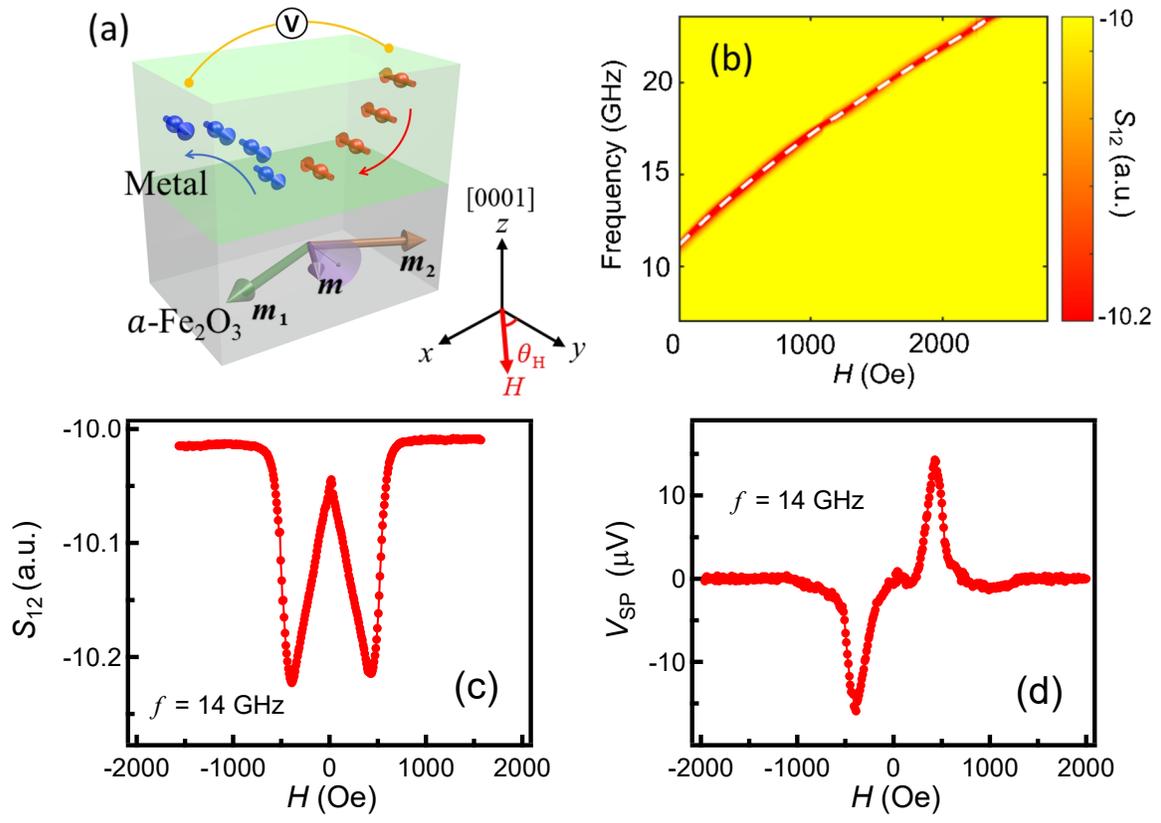

**Figure 2**

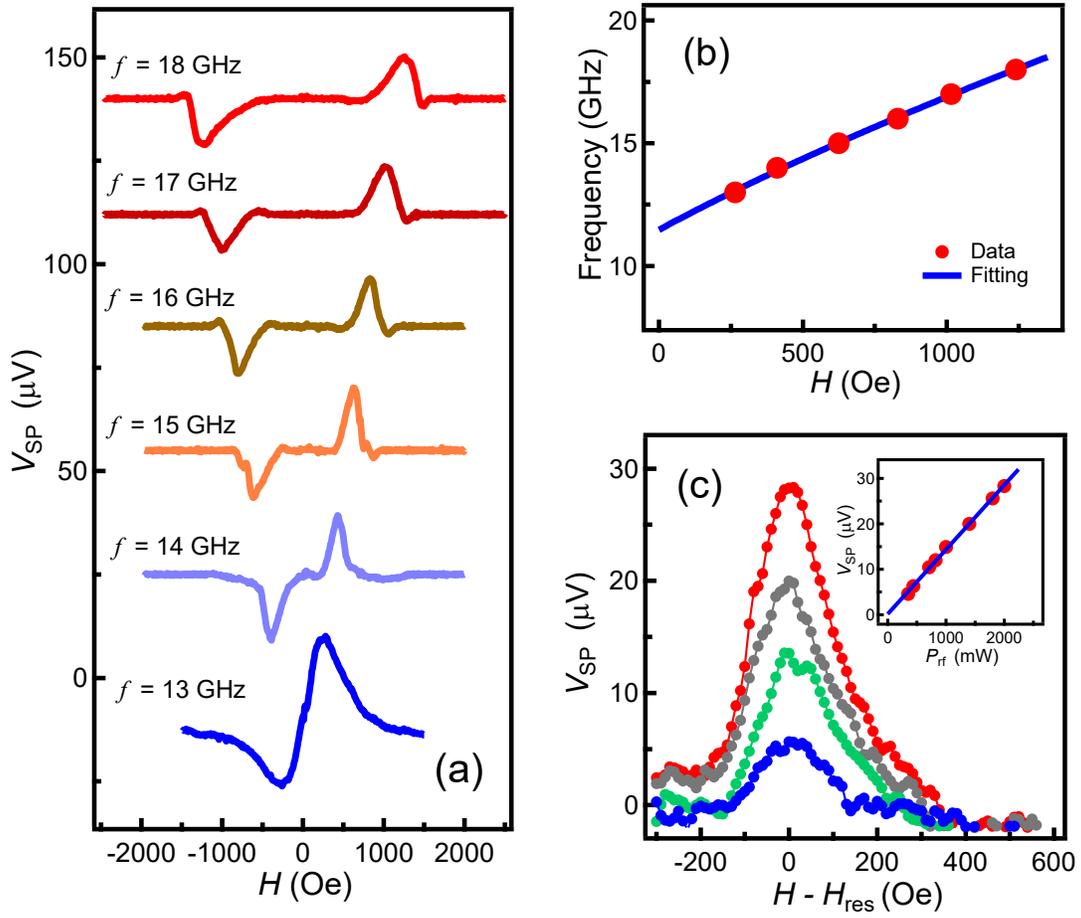

**Figure 3**

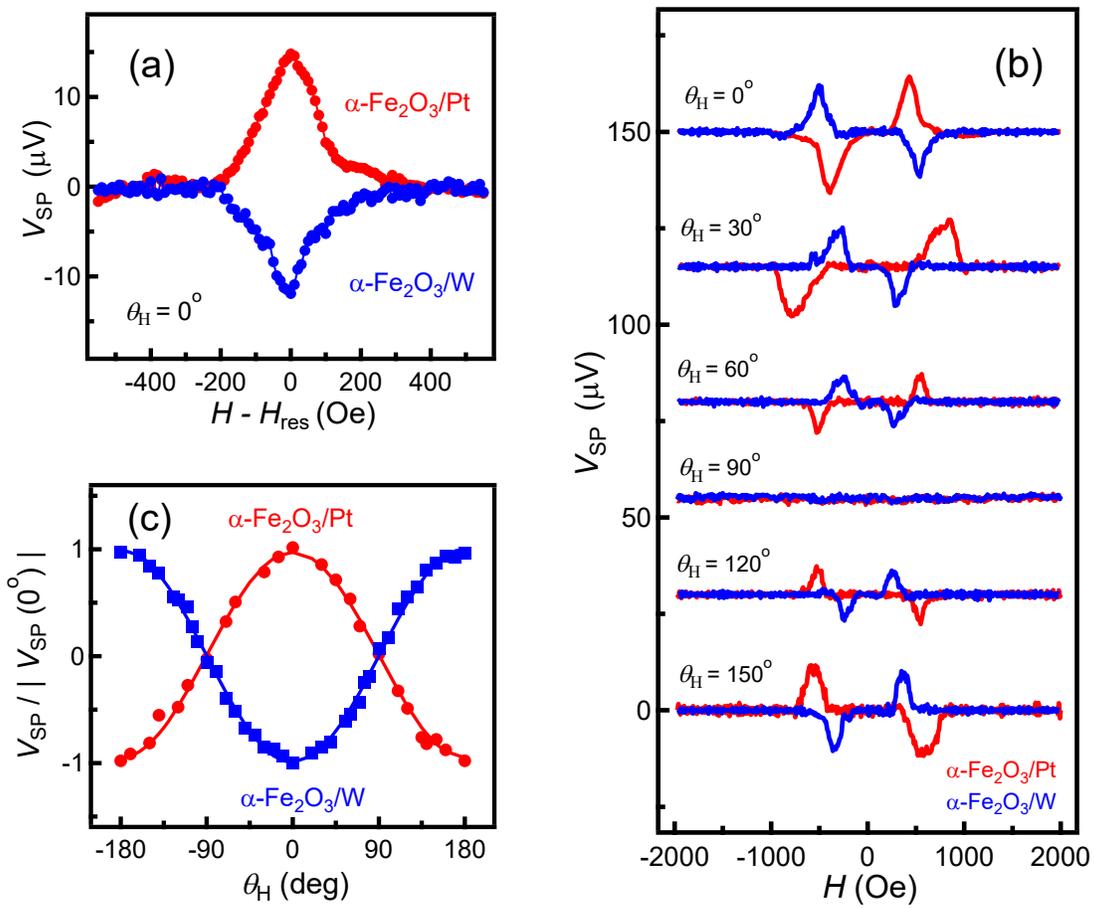

**Figure 4**

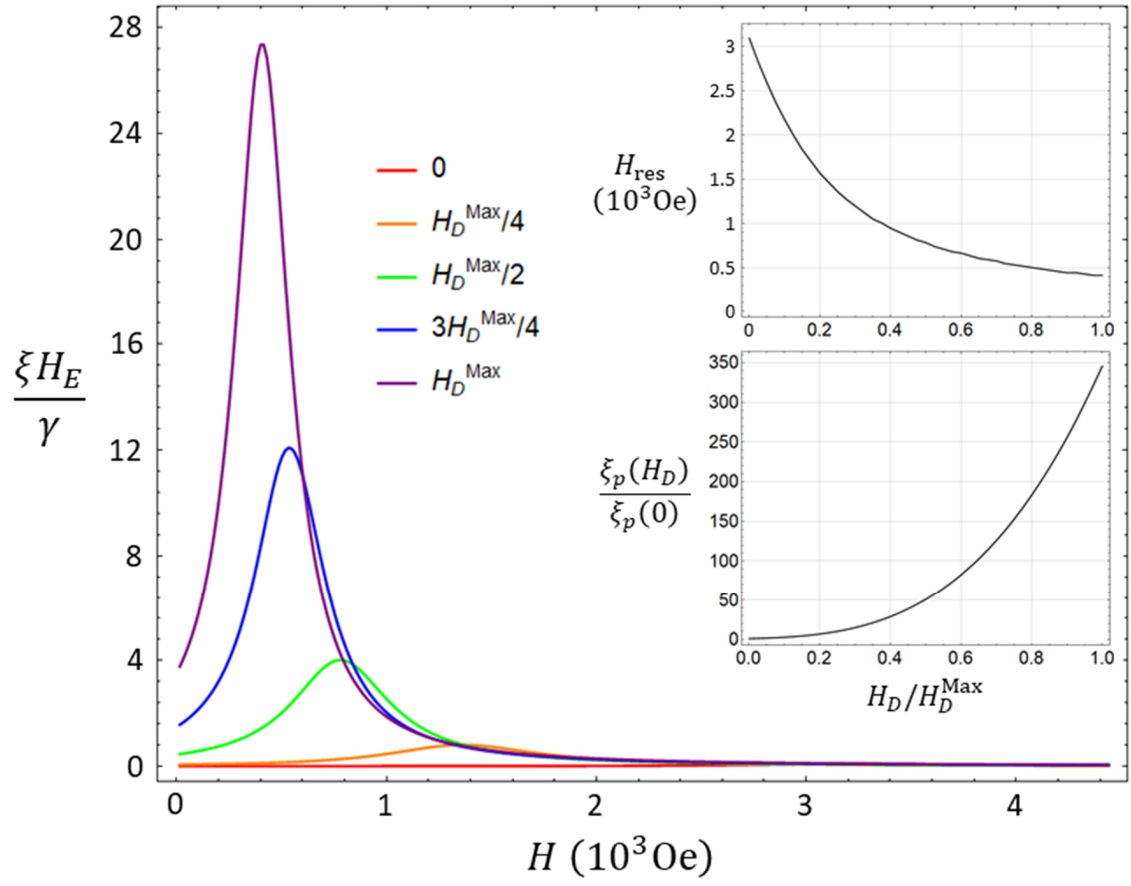

**Figure 5**